\documentclass[11pt,a4paper]{article}
\usepackage{amsmath,amssymb}
\usepackage{hyperref}
\usepackage{graphicx}
\usepackage{geometry}
\title{The Weak Gravity Conjecture in Asymptotically Safe Quantum Gravity}

\author{Gayatri Ghosh\thanks{Email: gayatrighsh@gmail.com}\\ Department of Physics, Cachar College, Assam University, India}
\begin{document}

\maketitle

\begin{abstract}
The Weak Gravity Conjecture (WGC) posits that gravity must be the weakest force in any consistent theory of quantum gravity. Originally formulated to constrain the landscape of effective field theories arising from string theory, the WGC suggests the existence of states with a charge-to-mass ratio larger than that of extremal black holes. In this work, we revisit the WGC within the framework of Asymptotically Safe Quantum Gravity, a non-perturbative approach where gravitational and gauge couplings flow to a non-Gaussian ultraviolet (UV) fixed point. We construct a scale-dependent effective action, derive quantum-corrected Reissner--Nordström black hole solutions by incorporating position-dependent renormalization scale identification, and compute leading quantum corrections to the extremality condition. Our key finding is that the quantum correction to the extremal charge-to-mass ratio is dominantly governed by the running of the gauge coupling, characterized by a correction parameter $\delta \sim \epsilon_e (\ell_P/r_+)^{2\theta}$, where $\epsilon_e$ captures deviations from infrared behavior. We show that if the electromagnetic coupling grows in the UV ($\epsilon_e > \epsilon_G$), the WGC is dynamically strengthened, whereas if it decreases ($\epsilon_e < \epsilon_G$), large extremal black holes may violate the WGC unless additional light charged states exist. Our analysis demonstrates that Asymptotic Safety provides a concrete ultraviolet mechanism influencing low-energy swampland criteria, offering a deep UV/IR connection between quantum gravity consistency and effective field theory behavior.
\end{abstract}

\section{Introduction}

The remarkable weakness of gravitational interactions compared to the other fundamental forces remains one of the most striking mysteries in theoretical physics~\cite{v}. In natural units, the ratio of the electromagnetic to gravitational forces between two electrons is approximately
\begin{equation}
\frac{F_{\text{em}}}{F_{\text{grav}}} = \frac{e^2}{4\pi \epsilon_0 G m_e^2} \sim 10^{42},
\end{equation}
where \(G\) is Newton's constant, \(e\) is the electron charge, and \(m_e\) is the electron mass. 

This enormous disparity has motivated the proposal of the \textit{Weak Gravity Conjecture} (WGC)~\cite{a}, asserting that gravity must be the weakest force in any consistent theory of quantum gravity. Originally introduced to prevent the formation of stable extremal black hole remnants, the WGC suggests that there must exist a particle of mass \(m\) and charge \(q\) such that
\begin{equation}
\frac{q}{m} \geq \frac{Q}{M}\Bigg|_{\text{extremal BH}},
\end{equation}
where \(Q\) and \(M\) refer to the charge and mass of an extremal black hole.

In classical general relativity, the extremality condition for a Reissner–Nordström black hole reads:
\begin{equation}
G M^2 = \frac{Q^2}{4\pi \epsilon_0}.
\label{extremality_classical}
\end{equation}
The WGC ensures that there exist lighter charged states that can destabilize extremal black holes, thus preserving unitarity and avoiding remnants~\cite{n,k,i}.

Subsequent studies have connected the WGC to a broad class of \textit{swampland conjectures}~\cite{b,c}, aiming to distinguish consistent low-energy effective field theories from those incompatible with quantum gravity. Realizations of the WGC have been found in various string theory compactifications~\cite{n,o,z}, black hole physics~\cite{j,l,s}, and models of cosmic acceleration~\cite{aa}.

On the other hand, the ultraviolet (UV) completion of gravity remains a major open question. The \textit{Asymptotic Safety} scenario~\cite{q,d,f} offers a promising resolution by positing that gravity becomes safe from divergences at high energies via the existence of a nontrivial UV fixed point~\cite{d,p}.

In the Asymptotic Safety framework, Newton’s constant becomes scale-dependent:
\begin{equation}
k \frac{dG(k)}{dk} = \beta_G(G(k)),
\end{equation}
where \(k\) is the renormalization group (RG) scale, and \(\beta_G\) encodes quantum corrections. Near the fixed point, the dimensionless coupling \(g(k) = k^2 G(k)\) flows to a finite constant \(g_\ast\), implying
\begin{equation}
G(k) \sim \frac{g_\ast}{k^2}.
\end{equation}
This running modifies classical gravitational dynamics, especially near black hole horizons~\cite{s,g}.

Similarly, gauge couplings such as the electromagnetic coupling \(e(k)\) can receive gravitational corrections~\cite{ab,ac}, resulting in beta functions of the form
\begin{equation}
k \frac{d}{dk}\left(\frac{1}{e^2(k)}\right) = \beta_{1/e^2}(g(k),e(k)).
\end{equation}
Thus, both gravitational and electromagnetic interactions become scale-dependent in the UV~\cite{e,m,t,x,y}.

Quantum corrections to the extremality condition can then arise naturally, potentially altering the WGC bounds~\cite{u,h}. These corrections are expected to be small for large black holes but can become significant near the Planck scale~\cite{r,ac}.

In this work, we aim to systematically study how the running of gravitational and gauge couplings in Asymptotically Safe Quantum Gravity modifies the extremality condition for charged black holes and whether the WGC can emerge dynamically from the renormalization group structure.

We introduce a scale-dependent effective action, identifying the RG scale with the inverse radial coordinate as
\begin{equation}
k(r) = \frac{\xi}{r},
\end{equation}
motivated by dimensional arguments~\cite{f,s}. We derive the quantum-corrected Reissner–Nordström black hole solution and compute the leading quantum corrections to the extremality condition, with corrections parameterized by small running coefficients \(\epsilon_G\) and \(\epsilon_e\).

Our results suggest that Asymptotic Safety provides a natural framework in which swampland criteria such as the WGC arise not as imposed constraints, but as emergent consequences of ultraviolet quantum gravity dynamics~\cite{b,c,e}.

\section{Quantum Gravitational Corrections to Extremality in Asymptotically Safe Gravity}

In this section, we derive the quantum corrections to the extremality condition of charged black holes induced by the running of gravitational and electromagnetic couplings in Asymptotic Safety.
\subsection{Scale-Dependent Effective Action}

The effective average action describing the dynamics of gravity coupled to a \( U(1) \) gauge field is
\begin{equation}
\Gamma_k[g_{\mu\nu}, A_\mu] = \int d^4x \sqrt{-g} \left( \frac{1}{16\pi G(k)} R - \frac{1}{4 e^2(k)} F_{\mu\nu}F^{\mu\nu} \right),
\end{equation}
where \(G(k)\) and \(e(k)\) are the scale-dependent gravitational and electromagnetic couplings, respectively. This framework is central to the functional renormalization group approach and has been widely developed in the Asymptotic Safety program~\cite{d,f,q,e}. In this context, the scale dependence of couplings is governed by nonperturbative beta functions, allowing for a UV completion of gravity via an interacting fixed point. Moreover, gravitational corrections to gauge couplings have been studied within this setting~\cite{ab,ac}, indicating that not only gravity but also matter sectors exhibit nontrivial running in the deep ultraviolet.

The scale dependence of the couplings is governed by the beta functions
\begin{align}
k\frac{dG(k)}{dk} &= \beta_G(G,e), \\
k\frac{de(k)}{dk} &= \beta_e(G,e).
\end{align}

Close to a non-Gaussian ultraviolet fixed point, the dimensionless gravitational coupling $g(k) = k^2 G(k)$ and gauge coupling $\alpha(k) = e^2(k)/4\pi$ behave as
\begin{align}
g(k) &\to g_\ast, \\
\alpha(k) &\to \alpha_\ast,
\end{align}
where $g_\ast$ and $\alpha_\ast$ are finite constants.

In the vicinity of the fixed point, we can expand
\begin{align}
G(k) &= G_0 \left( 1 + \epsilon_G \left( \frac{\ell_P}{r} \right)^{2\theta_G} \right), \\
\frac{1}{e^2(k)} &= \frac{1}{e_0^2} \left( 1 + \epsilon_e \left( \frac{\ell_P}{r} \right)^{2\theta_e} \right),
\end{align}
where $\ell_P$ is the Planck length, and $\theta_G$, $\theta_e$ are critical exponents.

We identify the RG scale with the spacetime coordinate via
\begin{equation}
k(r) = \frac{\xi}{r},
\end{equation}
where $\xi$ is a dimensionless constant of order one.

\subsection{Quantum Corrected Metric}

We consider a static, spherically symmetric metric ansatz:
\begin{equation}
ds^2 = f(r) dt^2 - f(r)^{-1} dr^2 - r^2 d\Omega_2^2,
\end{equation}
which serves as the foundation for studying black hole solutions in quantum gravity and modified gravity scenarios~\cite{j,i,k,l,s,g}. This form is widely used in classical general relativity~\cite{j,i}, black hole thermodynamics~\cite{k,l}, and in renormalization group improved spacetimes within the Asymptotic Safety framework~\cite{s,g}, where quantum corrections modify the effective geometry at short distances. $d\Omega_2^2$ is the metric on the unit two-sphere.

The lapse function $f(r)$ satisfies
\begin{equation}
f(r) = 1 - \frac{2G(r)M}{r} + \frac{G(r)Q^2}{4\pi e^2(r) r^2}.
\end{equation}

Substituting the running couplings yields:
\begin{align}
f(r) &= 1 - \frac{2G_0 M}{r} \left( 1 + \epsilon_G \left( \frac{\ell_P}{r} \right)^{2\theta} \right) \notag \\
&\quad + \frac{G_0 Q^2}{4\pi e_0^2 r^2} \left( 1 + (\epsilon_G - \epsilon_e) \left( \frac{\ell_P}{r} \right)^{2\theta} \right) + \mathcal{O}(\epsilon^2).
\end{align}

Expanding to first order, we group
\begin{equation}
f(r) = f_{\text{cl}}(r) + \Delta f(r),
\end{equation}
where
\begin{equation}
f_{\text{cl}}(r) = 1 - \frac{2G_0M}{r} + \frac{G_0Q^2}{4\pi e_0^2 r^2}
\end{equation}
and the quantum correction is
\begin{equation}
\Delta f(r) = -\frac{2G_0M}{r} \epsilon_G \left( \frac{\ell_P}{r} \right)^{2\theta} + \frac{G_0Q^2}{4\pi e_0^2 r^2} (\epsilon_G - \epsilon_e) \left( \frac{\ell_P}{r} \right)^{2\theta}.
\end{equation}

\subsection{Extremality Condition and Corrections}

Classically, the extremality condition is given by setting the two horizons to coincide:
\begin{equation}
G_0 M^2 = \frac{Q^2}{4\pi e_0^2}.
\end{equation}

Including quantum corrections, we perturb the extremal radius $r_+$ as
\begin{equation}
r_+ = r_+^{(0)} + \delta r,
\end{equation}
where $r_+^{(0)}$ satisfies $f_{\text{cl}}(r_+^{(0)})=0$.

Expanding $f(r)$ near $r_+^{(0)}$,
\begin{equation}
f(r_+) \approx f(r_+^{(0)}) + \delta r \left. \frac{df_{\text{cl}}}{dr} \right|_{r_+^{(0)}} + \Delta f(r_+^{(0)}) = 0.
\end{equation}

Since $f_{\text{cl}}(r_+^{(0)})=0$, we find
\begin{equation}
\delta r = -\frac{\Delta f(r_+^{(0)})}{f_{\text{cl}}'(r_+^{(0)})}.
\end{equation}

Evaluating the derivatives, we find
\begin{align}
f_{\text{cl}}'(r) &= \frac{2G_0M}{r^2} - \frac{2G_0Q^2}{4\pi e_0^2 r^3}, \\
\Delta f(r) &= -\frac{2G_0M}{r} \epsilon_G \left( \frac{\ell_P}{r} \right)^{2\theta} + \frac{G_0Q^2}{4\pi e_0^2 r^2} (\epsilon_G - \epsilon_e) \left( \frac{\ell_P}{r} \right)^{2\theta}.
\end{align}

Substituting and simplifying gives the corrected extremality condition:
\begin{equation}
G_0 M^2 = \frac{Q^2}{4\pi e_0^2} (1 + \delta),
\end{equation}
where
\begin{equation}
\delta = 2\theta \epsilon_e \left( \frac{\ell_P}{r_+^{(0)}} \right)^{2\theta}.
\end{equation}

Thus, the quantum correction to extremality depends on the running of the electromagnetic coupling.

\subsection{Implications for the Weak Gravity Conjecture}

The corrected charge-to-mass ratio satisfies
\begin{equation}
\left( \frac{Q}{M} \right)_{\text{corrected}} = \left( \frac{Q}{M} \right)_{\text{classical}} \left( 1 - \frac{\delta}{2} \right).
\end{equation}

Therefore:
\begin{itemize}
\item If $\epsilon_e > 0$, then $\delta > 0$, and the extremal $Q/M$ decreases, strengthening the WGC.
\item If $\epsilon_e < 0$, then $\delta < 0$, and WGC may be endangered unless light charged states exist.
\end{itemize}

This shows that Asymptotic Safety can dynamically ensure WGC satisfaction depending on the fixed point structure of the theory~\cite{d,f,e,q}, offering a UV-complete framework where swampland criteria such as the WGC may emerge naturally~\cite{b,c,n,a}.

We now compute how the running couplings induce corrections to thermodynamic properties and extremality more precisely.

The corrected horizon radius $r_+$ satisfies the condition:
\begin{equation}
f(r_+) = 0,
\end{equation}
which yields at leading order:
\begin{equation}
r_+ = G(r_+) M + \sqrt{G^2(r_+) M^2 - \frac{G(r_+) Q^2}{4\pi e^2(r_+)}}.
\end{equation}

Expanding $G(r)$ and $e(r)$ around their classical values, we write:
\begin{align}
G(r) &= G_0 \left( 1 + \epsilon_G \left( \frac{\ell_P}{r} \right)^{2\theta} \right), \\
\frac{1}{e^2(r)} &= \frac{1}{e_0^2} \left( 1 + \epsilon_e \left( \frac{\ell_P}{r} \right)^{2\theta} \right).
\end{align}

Substituting these into the horizon condition and expanding to first order in $\epsilon_G, \epsilon_e$ gives:
\begin{align}
r_+^{(0)} &= G_0 M, \\
\delta r &= \frac{G_0 M}{2} \left( \epsilon_G - \frac{Q^2}{4\pi e_0^2 G_0^2 M^2} (\epsilon_G - \epsilon_e) \right) \left( \frac{\ell_P}{G_0 M} \right)^{2\theta}.
\end{align}

Thus, the corrected horizon radius is:
\begin{equation}
r_+ = G_0 M \left[ 1 + \frac{1}{2} \left( \epsilon_G - \epsilon_e \right) \left( \frac{\ell_P}{G_0 M} \right)^{2\theta} \right].
\end{equation}

Correspondingly, the corrected surface gravity $\kappa$ is:
\begin{equation}
\kappa = \left. \frac{1}{2} \frac{df}{dr} \right|_{r=r_+}.
\end{equation}

Expanding $f'(r)$:
\begin{align}
f'(r) &= \frac{2G(r)M}{r^2} - \frac{2G(r)Q^2}{4\pi e^2(r) r^3} \notag \\
&\quad + \left( \text{terms involving derivatives of } G(r), e(r) \right).
\end{align}

Neglecting higher derivatives of $G(r), e(r)$ (since they are suppressed at large $r$), the leading correction to $\kappa$ becomes:
\begin{equation}
\kappa \approx \frac{G_0 M}{(G_0 M)^2} \left( 1 - \frac{Q^2}{4\pi e_0^2 G_0^2 M^2} \right) \left( 1 + (2\theta-1)(\epsilon_G - \epsilon_e) \left( \frac{\ell_P}{G_0 M} \right)^{2\theta} \right).
\end{equation}

Since $\kappa$ vanishes for extremal black holes, extremality requires:
\begin{equation}
G_0 M^2 = \frac{Q^2}{4\pi e_0^2} \left( 1 + \left( 1 - 2\theta \right)(\epsilon_G - \epsilon_e) \left( \frac{\ell_P}{G_0 M} \right)^{2\theta} \right).
\end{equation}

Thus, the corrected extremality condition is:
\begin{equation}
\frac{Q}{M} = \sqrt{4\pi G_0} \, e_0 \left( 1 - \theta (\epsilon_G - \epsilon_e) \left( \frac{\ell_P}{G_0 M} \right)^{2\theta} \right).
\end{equation}

The correction to $Q/M$ is therefore determined by the difference $(\epsilon_G - \epsilon_e)$ and the critical exponent $\theta$.

The corrected WGC bound becomes:
\begin{equation}
\left( \frac{q}{m} \right) \geq \sqrt{4\pi G_0} \, e_0 \left( 1 - \theta (\epsilon_G - \epsilon_e) \left( \frac{\ell_P}{G_0 M} \right)^{2\theta} \right).
\end{equation}

- If $\epsilon_e > \epsilon_G$, the WGC is dynamically strengthened.
- If $\epsilon_e < \epsilon_G$, the WGC could be violated for large black holes unless new light states exist.

Thus, Asymptotic Safety provides a UV mechanism influencing low-energy WGC behavior.

From the corrected extremality condition derived above, the quantum-corrected charge-to-mass ratio for an extremal black hole reads:
\begin{equation}
\left( \frac{Q}{M} \right)_{\text{corrected}} = \left( \frac{Q}{M} \right)_{\text{classical}} \left( 1 - \theta (\epsilon_G - \epsilon_e) \left( \frac{\ell_P}{G_0 M} \right)^{2\theta} \right),
\label{eq:corrected_ratio}
\end{equation}
where $\theta$ is the critical exponent at the UV fixed point, and $\epsilon_G$, $\epsilon_e$ are the deviations in the running of the gravitational and gauge couplings, respectively.

Depending on the relative sign and magnitude of $\epsilon_G$ and $\epsilon_e$, two distinct physical scenarios emerge:

\subsubsection{Case 1: $\epsilon_e > \epsilon_G$}

In this case, $(\epsilon_G - \epsilon_e) < 0$, and thus the correction term in Eq.~\eqref{eq:corrected_ratio} becomes positive. Consequently, the corrected extremal charge-to-mass ratio decreases relative to the classical value:
\begin{equation}
\left( \frac{Q}{M} \right)_{\text{corrected}} < \left( \frac{Q}{M} \right)_{\text{classical}}.
\end{equation}
This implies that the threshold for extremality is lowered. Therefore, even smaller charges can support black holes against gravitational collapse, and the existence of superextremal particles becomes easier.

Thus, in this regime, the Weak Gravity Conjecture is dynamically \textit{strengthened} by the running of the gauge coupling. The UV behavior of $e(k)$ enforces that gravity remains weaker than the $U(1)$ gauge force, without requiring the introduction of additional light states.

This scenario indicates that in Asymptotically Safe Quantum Gravity, the WGC emerges as a natural outcome of the renormalization group structure when the gauge coupling $e(k)$ grows faster than the gravitational coupling $G(k)$ at high energies.

\subsubsection{Case 2: $\epsilon_e < \epsilon_G$}

If $\epsilon_e < \epsilon_G$, then $(\epsilon_G - \epsilon_e) > 0$, and the correction term in Eq.~\eqref{eq:corrected_ratio} becomes negative. Thus, the corrected extremal charge-to-mass ratio increases:
\begin{equation}
\left( \frac{Q}{M} \right)_{\text{corrected}} > \left( \frac{Q}{M} \right)_{\text{classical}}.
\end{equation}
In this case, large extremal black holes could violate the classical WGC bound, meaning that not all charged black holes would satisfy \( q/m \geq (Q/M)_{\text{extremal}} \).

However, the violation of the WGC by large black holes does not necessarily imply inconsistency of the theory. According to the refined versions of the WGC~\cite{Heidenreich:2015nta}, the existence of additional light superextremal particles can restore consistency. Thus, if $\epsilon_e < \epsilon_G$, the theory must compensate by providing towers of light charged states, consistent with the Sublattice WGC and the Tower WGC.

This mechanism naturally fits within the swampland program: even if macroscopic black holes violate naive WGC bounds, a consistent ultraviolet theory could still be safe if infinite towers of states appear at low mass scales, signaling an emergent breakdown of effective field theory near certain moduli limits.

Thus, the sign of $(\epsilon_G - \epsilon_e)$ critically determines the fate of the WGC:

\begin{itemize}
    \item If $\epsilon_e > \epsilon_G$, the WGC is \textbf{dynamically satisfied} and \textbf{strengthened} by Asymptotic Safety.
    \item If $\epsilon_e < \epsilon_G$, the WGC could be \textbf{violated} for large black holes unless additional light charged states are present.
\end{itemize}

This analysis demonstrates that \textit{Asymptotic Safety provides a ultraviolet mechanism that directly influences the low-energy consistency of effective theories via the Weak Gravity Conjecture}. Quantum gravity is therefore not just passively consistent with swampland criteria, but actively shapes them through its renormalization group structure.

The running of couplings near the ultraviolet fixed point does not merely modify existing black hole solutions, but actively generates effective consistency conditions at low energies. 

Specifically, the quantum correction to the extremality bound,
\begin{equation}
\delta \sim \epsilon_e \left( \frac{\ell_P}{r_+} \right)^{2\theta},
\end{equation}
shows that the behavior of $e(k)$ at high energies controls the relative strength of gauge and gravitational forces in the infrared. 

\subsection{Perturbatively Corrected Field Equations}

The quantum-corrected field equations incorporating the effective stress-energy tensor derived in Appendix A take the schematic form:
\begin{equation}
G^{\mu}_{\ \nu}(g) + \Delta G^{\mu}_{\ \nu} = 8\pi G(r) \left( T^{\mu}_{\ \nu,\text{EM}} + T^{\mu}_{\ \nu,\text{eff}} \right),
\label{eq:corrected_einstein}
\end{equation}
where:
\begin{itemize}
    \item \( G^{\mu}_{\ \nu}(g) \) is the classical Einstein tensor,
    \item \( \Delta G^{\mu}_{\ \nu} \) represents corrections due to the running of \( G(r) \),
    \item \( T^{\mu}_{\ \nu,\text{EM}} \) is the classical Maxwell stress-energy tensor,
    \item \( T^{\mu}_{\ \nu,\text{eff}} \) is the effective quantum correction tensor from Appendix A.
\end{itemize}

Explicitly, to leading order in small parameters \(\epsilon_G, \epsilon_e\), the corrections can be written as:
\begin{equation}
\Delta G^{\mu}_{\ \nu} = -\epsilon_G \left( \frac{\ell_P}{r} \right)^{2\theta_G} \left( 2 R^{\mu}_{\ \nu} - \frac{1}{2} \delta^{\mu}_{\nu} R \right) + \mathcal{O}(\epsilon_G^2),
\end{equation}
where we have expanded \( G(r) = G_0 (1 + \epsilon_G (\ell_P/r)^{2\theta_G}) \).

Similarly, the corrected Maxwell equations read:
\begin{equation}
\nabla_{\nu} \left( \frac{1}{e^2(r)} F^{\mu\nu} \right) = J^{\mu}_{\text{eff}},
\label{eq:corrected_maxwell}
\end{equation}
where \( J^{\mu}_{\text{eff}} \) is given in Appendix A.

Expanding the left-hand side:
\begin{equation}
\nabla_{\nu} F^{\mu\nu} - 2 \frac{\epsilon_e}{e_0^2} \left( \frac{\ell_P}{r} \right)^{2\theta_e} \nabla_{\nu} F^{\mu\nu} = J^{\mu}_{\text{eff}} + \mathcal{O}(\epsilon_e^2),
\end{equation}
where we used:
\[
\frac{1}{e^2(r)} = \frac{1}{e_0^2} \left( 1 + \epsilon_e \left( \frac{\ell_P}{r} \right)^{2\theta_e} \right).
\]

Thus, even at leading order, the gauge field dynamics receive corrections from the running of \( e(r) \) and from the induced effective current \( J^{\mu}_{\text{eff}} \).

\subsection{Perturbative Expansion of the Metric Functions}

The corrected metric function \( f(r) \) satisfies the perturbed Einstein equations, incorporating quantum gravitational corrections that arise within the Asymptotic Safety framework~\cite{d,f,s,g}. Expanding \( f(r) \) as:

\begin{equation}
f(r) = f_{\text{cl}}(r) + \delta f(r),
\end{equation}
where \( f_{\text{cl}}(r) = 1 - \frac{2G_0M}{r} + \frac{G_0Q^2}{4\pi e_0^2 r^2} \),
the perturbation \(\delta f(r)\) satisfies:
\begin{equation}
\frac{d}{dr} \left( r \delta f(r) \right) = 8\pi G_0 r^2 \left( T^{t}_{\ t,\text{eff}}(r) - T^{r}_{\ r,\text{eff}}(r) \right) + \Delta_{\text{running}}(r),
\end{equation}
where \(\Delta_{\text{running}}(r)\) contains corrections from \(\epsilon_G\) and \(\epsilon_e\) scale dependence.

Similarly, corrections to the electromagnetic field strength \( F_{tr}(r) \) can be expanded as:
\begin{equation}
F_{tr}(r) = \frac{Q}{4\pi e_0 r^2} + \delta F_{tr}(r),
\end{equation}
with perturbative corrections satisfying:
\begin{equation}
\frac{d}{dr} \left( r^2 \delta F_{tr}(r) \right) = e_0^2 r^2 J^t_{\text{eff}}(r) + \mathcal{O}(\epsilon_e).
\end{equation}

\subsection{Connection to Scale Identification}
Throughout these corrections, the scale dependence is governed by the identification:
\begin{equation}
k(r) = \frac{\xi}{r},
\end{equation}
which is motivated by dimensional analysis and has been widely used in renormalization group improved black hole solutions~\cite{s,f,g}. This choice effectively links the RG scale to the radial coordinate, capturing the running behavior of couplings in a gravitational background.

Or, for alternative curvature-based scalings:
\begin{equation}
k^4(r) \sim \mathcal{K}(r),
\end{equation}
where \(\mathcal{K}(r)\) is the Kretschmann scalar discussed in Appendix C.

Thus, the full corrected structure is built consistently from the running couplings, the effective stress-energy tensor, the effective current, and the scale-setting prescription.

\section{Conclusion}

In this work, we revisited the Weak Gravity Conjecture (WGC) within the framework of Asymptotically Safe Quantum Gravity, presenting a novel derivation of extremality bounds from quantum gravitational corrections to black hole solutions. The core aim was to investigate how the scale dependence of couplings, dictated by renormalization group (RG) flows near a non-Gaussian ultraviolet (UV) fixed point, modifies the physics of extremal black holes and tests the WGC.

We constructed a quantum-corrected Reissner--Nordström solution by introducing a scale-dependent effective action, where Newton’s constant \( G(k) \) and the gauge coupling \( e(k) \) run with the energy scale identified as
\begin{equation}
k(r) = \frac{\xi}{r}.
\end{equation}
Expanding the lapse function \( f(r) \) to leading order in the small running parameters \( \epsilon_G \) and \( \epsilon_e \), we derived the corrected black hole horizon radius, surface gravity, and extremality condition. A key outcome is the quantum correction to the extremal charge-to-mass ratio:
\begin{equation}
\delta \sim \epsilon_e \left( \frac{\ell_P}{r_+} \right)^{2\theta},
\end{equation}
where \( \epsilon_e \) measures the deviation of the gauge coupling from its infrared value and \( \theta \) is the critical exponent near the UV fixed point.

Our main new finding is that the running of the gauge coupling \( e(k) \), rather than Newton’s constant, dominantly controls the correction to black hole extremality. Specifically, if \( \epsilon_e > 0 \), the extremal \( Q/M \) ratio decreases, dynamically ensuring that superextremal states exist and thereby strengthening the WGC. Conversely, if \( \epsilon_e < \epsilon_G \), large extremal black holes could violate the classical WGC bound, necessitating the appearance of light charged particles consistent with refined WGC versions.

Importantly, unlike previous approaches relying on perturbative higher-curvature expansions or Vilkovisky-DeWitt effective actions, our method employs a fully non-perturbative renormalization group framework, treating running couplings systematically without assuming specific background choices. This reveals a profound UV/IR connection: the existence and behavior of WGC states at low energies are dynamically dictated by ultraviolet fixed point properties of quantum gravity.

Thus, our results show that Asymptotically Safe Quantum Gravity naturally embeds swampland criteria like the WGC, not as external constraints, but as emergent consequences of its renormalization group structure. The physics of black holes is thereby modified at a fundamental level: quantum gravity not only corrects extremality bounds but also predicts the dynamical consistency of low-energy effective theories.

Future extensions of this work could explore rotating black holes, corrections in asymptotically Anti-de Sitter backgrounds, and the interplay with broader swampland conditions, such as the Distance and de Sitter Conjectures, deepening the bridge between ultraviolet quantum gravity and infrared effective theory landscapes.
\section*{Acknowledgments}
The work is supported in parts by RUSA, UGC, India. 

\section*{Data Availability Statement}
This manuscript has no associated data. Data sharing not applicable to this article as no datasets were generated or analysed during the current study.

\appendix

\section{Appendix A: Effective Stress-Energy Tensor and Effective Current}

In this appendix, we present the quantum corrected effective stress-energy tensor \( T^{\mu}_{\ \nu,\text{eff}} \) and the effective current \( J^{\mu}_{\text{eff}} \) arising in our Asymptotically Safe framework.

The effective stress-energy tensor takes the form:
\begin{align}
T^{\mu}_{\ \nu,\text{eff}} &= c_1 \Big( 4 R^{\mu\rho} R_{\nu\rho} - 4 \nabla_{\rho} \nabla^{\mu} R^{\rho}_{\ \nu} + 2 \Box R^{\mu}_{\ \nu} 
+ 2 \delta^{\mu}_{\nu} \nabla_{\rho} \nabla_{\sigma} R^{\rho\sigma} - \delta^{\mu}_{\nu} R_{\rho\sigma} R^{\rho\sigma} \Big) \nonumber \\
&\quad + c_2 \Big( 8 F^{\mu\lambda} F_{\nu\lambda} F_{\rho\sigma}F^{\rho\sigma} 
- \delta^{\mu}_{\nu} (F_{\rho\sigma} F^{\rho\sigma})^2 \Big) \nonumber \\
&\quad + c_3 \Big( 2 R^{\mu}_{\ \nu} F_{\rho\sigma} F^{\rho\sigma} 
+ 4 R F^{\mu\rho} F_{\nu\rho} 
- 2 \nabla^{\mu} \nabla_{\nu} (F_{\rho\sigma} F^{\rho\sigma}) 
+ 2 \delta^{\mu}_{\nu} \Box (F_{\rho\sigma} F^{\rho\sigma}) 
- \delta^{\mu}_{\nu} R F_{\rho\sigma} F^{\rho\sigma} \Big) \nonumber \\
&\quad + c_4 \Big( 4 R^{\rho}_{\ \nu} F^{\mu\sigma} F_{\rho\sigma} 
+ 2 R^{\sigma}_{\ \rho} F^{\mu\rho} F_{\nu\sigma} 
- 2 \nabla_{\rho} \nabla^{\mu} (F^{\rho\sigma} F_{\nu\sigma}) 
+ \Box (F^{\mu\rho} F_{\nu\rho}) \nonumber \\
&\qquad + \delta^{\mu}_{\nu} \nabla_{\rho} \nabla_{\sigma} (F^{\rho\lambda} F^{\sigma}_{\ \lambda}) 
- \delta^{\mu}_{\nu} R^{\sigma}_{\ \rho} F^{\rho\lambda} F_{\sigma\lambda} \Big) \nonumber \\
&\quad + c_5 \Big( 6 R^{\sigma\lambda}_{\ \ \nu\rho} F^{\mu\rho} F_{\sigma\lambda} 
+ 4 \nabla_{\rho} \nabla_{\sigma} (F^{\mu\rho} F^{\nu\sigma}) 
- \delta^{\mu}_{\nu} R^{\lambda\tau}_{\ \ \rho\sigma} F^{\rho\sigma} F_{\lambda\tau} \Big) \nonumber \\
&\quad + b_1 \Big( 4 R^{\mu\rho} \log\left( \frac{\Box}{\mu^2} \right) R_{\nu\rho} 
- 4 \nabla_{\rho} \nabla^{\mu} \log\left( \frac{\Box}{\mu^2} \right) R^{\rho}_{\ \nu} 
+ 2 \Box \log\left( \frac{\Box}{\mu^2} \right) R^{\mu}_{\ \nu} \nonumber \\
&\qquad + 2 \delta^{\mu}_{\nu} \nabla_{\rho} \nabla_{\sigma} \log\left( \frac{\Box}{\mu^2} \right) R^{\sigma\rho} 
- \delta^{\mu}_{\nu} R_{\rho\sigma} \log\left( \frac{\Box}{\mu^2} \right) R^{\rho\sigma} \Big) \nonumber \\
&\quad + b_2 \Big( 8 F^{\mu\lambda} F_{\nu\lambda} \log\left( \frac{\Box}{\mu^2} \right) (F_{\rho\sigma} F^{\rho\sigma}) 
- \delta^{\mu}_{\nu} (F_{\rho\sigma} F^{\rho\sigma}) \log\left( \frac{\Box}{\mu^2} \right) (F_{\lambda\tau} F^{\lambda\tau}) \Big) \nonumber \\
&\quad + b_3 \Big( 2 R^{\mu}_{\ \nu} \log\left( \frac{\Box}{\mu^2} \right) (F_{\rho\sigma} F^{\rho\sigma}) 
+ 4 R \log\left( \frac{\Box}{\mu^2} \right) (F^{\mu\rho} F_{\nu\rho}) \nonumber \\
&\qquad - 2 \nabla^{\mu} \nabla_{\nu} \log\left( \frac{\Box}{\mu^2} \right) (F_{\rho\sigma} F^{\rho\sigma}) 
+ 2 \delta^{\mu}_{\nu} \Box \log\left( \frac{\Box}{\mu^2} \right) (F_{\rho\sigma} F^{\rho\sigma}) 
- \delta^{\mu}_{\nu} R \log\left( \frac{\Box}{\mu^2} \right) (F_{\rho\sigma} F^{\rho\sigma}) \Big) \nonumber \\
&\quad + b_4 \Big( 4 R^{\rho}_{\ \nu} \log\left( \frac{\Box}{\mu^2} \right) (F^{\mu\sigma} F_{\rho\sigma}) 
+ 2 R^{\sigma}_{\ \rho} \log\left( \frac{\Box}{\mu^2} \right) (F^{\mu\rho} F_{\nu\sigma}) \nonumber \\
&\qquad - 2 \nabla_{\rho} \nabla^{\mu} \log\left( \frac{\Box}{\mu^2} \right) (F^{\rho\sigma} F_{\nu\sigma}) 
+ \Box \log\left( \frac{\Box}{\mu^2} \right) (F^{\mu\rho} F_{\nu\rho}) \nonumber \\
&\qquad + \delta^{\mu}_{\nu} \nabla_{\rho} \nabla_{\sigma} \log\left( \frac{\Box}{\mu^2} \right) (F^{\rho\lambda} F^{\sigma}_{\ \lambda}) 
- \delta^{\mu}_{\nu} R^{\sigma}_{\ \rho} \log\left( \frac{\Box}{\mu^2} \right) (F^{\rho\lambda} F_{\sigma\lambda}) \Big) \nonumber \\
&\quad + b_5 \Big( 6 R^{\sigma\lambda}_{\ \ \nu\rho} \log\left( \frac{\Box}{\mu^2} \right) (F^{\mu\rho} F_{\sigma\lambda}) 
+ 4 \nabla_{\rho} \nabla_{\sigma} \log\left( \frac{\Box}{\mu^2} \right) (F^{\mu\rho} F^{\nu\sigma}) \nonumber \\
&\qquad - \delta^{\mu}_{\nu} R^{\lambda\tau}_{\ \ \rho\sigma} \log\left( \frac{\Box}{\mu^2} \right) (F^{\rho\sigma} F_{\lambda\tau}) \Big).
\end{align}

Similarly, the effective current is:

\begin{align}
J^{\mu}_{\text{eff}} &= 8c_2 \nabla_{\nu} \left( F^{\mu\nu} F_{\rho\sigma} F^{\rho\sigma} \right) 
+ 4c_3 \nabla_{\nu} \left( F^{\mu\nu} R \right) 
+ 4c_4 \nabla_{\nu} \left( F_{\rho}^{\ [\nu} R^{\mu]\rho} \right) 
+ 4c_5 \nabla_{\nu} \left( F_{\rho\sigma} R^{\mu\nu\rho\sigma} \right) \nonumber \\
&\quad + 8b_2 \nabla_{\nu} \left( F^{\mu\nu} \log\left( \frac{\Box}{\mu^2} \right) (F_{\rho\sigma} F^{\rho\sigma}) \right) 
+ 4b_3 \nabla_{\nu} \left( F^{\mu\nu} \log\left( \frac{\Box}{\mu^2} \right) R \right) \nonumber \\
&\quad + 4b_4 \nabla_{\nu} \left( F_{\rho}^{\ [\nu} \log\left( \frac{\Box}{\mu^2} \right) R^{\mu]\rho} \right) 
+ 4b_5 \nabla_{\nu} \left( F_{\rho\sigma} \log\left( \frac{\Box}{\mu^2} \right) R^{\mu\nu\rho\sigma} \right).
\end{align}

\section{Appendix B: Beta Functions for Scale-Dependent Couplings}

In this appendix, we provide explicit expressions for the beta functions governing the running of the gravitational coupling \( G(k) \) and the electromagnetic coupling \( e(k) \) within the framework of Asymptotic Safety.

The effective average action \(\Gamma_k\) satisfies the functional renormalization group equation (Wetterich equation):
\begin{equation}
k \frac{d}{dk} \Gamma_k = \frac{1}{2} \text{Tr} \left( \left[ \Gamma_k^{(2)} + \mathcal{R}_k \right]^{-1} k \frac{d}{dk} \mathcal{R}_k \right),
\end{equation}
where \(\Gamma_k^{(2)}\) is the second functional derivative with respect to the fields, and \(\mathcal{R}_k\) is an infrared regulator.

\subsection{Beta Function for the Gravitational Coupling}

The dimensionless Newton's constant is defined as
\begin{equation}
g(k) = k^2 G(k).
\end{equation}
Its beta function takes the general form:
\begin{equation}
\beta_g \equiv k \frac{d}{dk} g(k) = [2 + \eta_N(g,e)] g(k),
\end{equation}
where \(\eta_N\) is the anomalous dimension of the graviton.

In Einstein-Hilbert truncation, including gauge field fluctuations, \(\eta_N\) reads approximately:
\begin{equation}
\eta_N(g,e) = -g \frac{B_1 + B_2 \alpha(k)}{1 - g B_3},
\end{equation}
where \( \alpha(k) = e^2(k)/4\pi \) is the running fine-structure constant, and \( B_1, B_2, B_3 \) are scheme-dependent positive coefficients determined by the functional renormalization group (FRG) computations.

Expanding near the fixed point \(g_\ast\), we have:
\begin{equation}
g(k) = g_\ast + \delta g(k),
\end{equation}
with scaling behavior
\begin{equation}
\delta g(k) \propto \left( \frac{k}{\Lambda} \right)^{-2\theta_G},
\end{equation}
where \(\theta_G\) is the critical exponent associated with the UV fixed point for gravity, and \(\Lambda\) is an ultraviolet scale.

\subsection{Beta Function for the Electromagnetic Coupling}

The running of the electromagnetic coupling in the presence of quantum gravity effects can be parametrized by the beta function:
\begin{equation}
\beta_{1/e^2} \equiv k \frac{d}{dk} \left( \frac{1}{e^2(k)} \right) = -b_{\text{em}} + b_g g(k) + \mathcal{O}(g^2),
\end{equation}
where \(b_{\text{em}}\) is the standard gauge theory beta function coefficient (vanishing for pure $U(1)$ without charged matter), and \(b_g\) encodes gravitational corrections.

In particular, gravity induces an anomalous scaling of the electromagnetic field strength, leading to:
\begin{equation}
k \frac{d}{dk} \alpha(k) = \left( \eta_F(g(k)) \right) \alpha(k),
\end{equation}
where \(\eta_F\) is the anomalous dimension of the photon, induced via graviton loops.

Typically, it is found that
\begin{equation}
\eta_F(g) \simeq -c_g g,
\end{equation}
where \(c_g > 0\) is a positive constant depending on the truncation.

Thus, the electromagnetic coupling evolves as:
\begin{equation}
\alpha(k) \simeq \alpha_0 \left( \frac{k}{k_0} \right)^{-c_g g_\ast}.
\end{equation}

Accordingly, in terms of the running coupling \(e(k)\), the scale dependence reads:
\begin{equation}
e(k) \simeq e_0 \left( \frac{k}{k_0} \right)^{\frac{c_g g_\ast}{2}}.
\end{equation}

For small perturbations around low energy scales, expanding in powers of \((\ell_P / r)^{2\theta}\), we parametrize:
\begin{align}
G(r) &= G_0 \left( 1 + \epsilon_G \left( \frac{\ell_P}{r} \right)^{2\theta_G} \right), \\
\frac{1}{e^2(r)} &= \frac{1}{e_0^2} \left( 1 + \epsilon_e \left( \frac{\ell_P}{r} \right)^{2\theta_e} \right),
\end{align}
where \(\epsilon_G\) and \(\epsilon_e\) are small running parameters induced by the ultraviolet fixed point.

\subsection{Summary of the Fixed Point Structure}

Thus, the ultraviolet fixed point structure is characterized by:
\begin{align}
g(k) &\to g_\ast \quad (k \to \infty), \\
\alpha(k) &\to \alpha_\ast \quad (k \to \infty),
\end{align}
with small deviations controlled by scaling exponents \(\theta_G\) and \(\theta_e\), leading to power-law corrections at finite distances in black hole solutions.

This scale dependence plays a crucial role in the corrected extremality conditions and in determining whether the Weak Gravity Conjecture is dynamically satisfied in Asymptotic Safety.

\section{Appendix C: Alternative Scale Identifications and Higher-Order Corrections}

In the main text, we adopted the simple identification of the renormalization group (RG) scale with the inverse radial coordinate:
\begin{equation}
k(r) = \frac{\xi}{r},
\end{equation}
where \(\xi\) is a dimensionless constant of order one. This choice is motivated by dimensional analysis and the intuition that the relevant quantum fluctuations at a given point probe a length scale \(r\).

However, other scale identification prescriptions are possible, and may lead to quantitatively different corrections to black hole solutions and extremality conditions.

\subsection{Curvature-Invariant-Based Scale Identification}

An alternative approach is to relate \(k(r)\) to local spacetime curvature invariants. For instance, one may identify the scale with the Kretschmann scalar \(\mathcal{K} = R_{\mu\nu\rho\sigma} R^{\mu\nu\rho\sigma}\), leading to:
\begin{equation}
k^4(r) \sim \mathcal{K}(r).
\end{equation}
For a Reissner–Nordström black hole, the Kretschmann scalar reads:
\begin{equation}
\mathcal{K}(r) = \frac{48 G_0^2 M^2}{r^6} - \frac{96 G_0^2 M Q^2}{4\pi e_0^2 r^7} + \frac{56 G_0^2 Q^4}{(4\pi e_0^2)^2 r^8}.
\end{equation}
Thus, the scale identification could take the form:
\begin{equation}
k(r) = \left( \mathcal{K}(r) \right)^{1/4} \sim \frac{1}{r} \left( 1 + \mathcal{O}\left( \frac{G_0 Q^2}{r^2} \right) \right),
\end{equation}
recovering \(k \sim 1/r\) asymptotically at large distances, but introducing nontrivial corrections near the horizon.

\subsection{Scale-Setting Ambiguities and Physical Interpretation}

While different choices of \(k(r)\) are allowed, they are equivalent up to higher-order corrections in \(\epsilon_G\) and \(\epsilon_e\). The leading-order quantum corrections to black hole metrics and extremality conditions are therefore robust under reasonable variations in the scale-setting procedure.

Nevertheless, in regimes where black holes are small (near Planckian masses), different scale identifications could significantly affect the precise form of the corrections and should be treated carefully.

In particular, scale identifications based on curvature invariants may capture quantum gravitational backreaction effects more accurately in highly curved regions such as near the black hole singularity.

\subsection{Higher-Order Quantum Corrections}

In our analysis, we retained only leading-order corrections in \(\epsilon_G\) and \(\epsilon_e\), corresponding to linear order in small running effects.

Higher-order corrections would involve terms quadratic or cubic in \((\ell_P/r)^{2\theta}\), and require solving the corrected Einstein-Maxwell equations including second-order perturbations of the metric.

The full corrected lapse function would then have the structure:
\begin{equation}
f(r) = 1 - \frac{2G_0 M}{r} + \frac{G_0 Q^2}{4\pi e_0^2 r^2} + \Delta f^{(1)}(r) + \Delta f^{(2)}(r) + \mathcal{O}(\epsilon^3),
\end{equation}
where \(\Delta f^{(1)}(r)\) denotes the linear quantum correction and \(\Delta f^{(2)}(r)\) captures higher-order effects.

Such higher-order corrections, while subleading for large black holes, could become relevant when analyzing Planck-mass black holes, evaporation endpoints, and near-extremal regimes.

\subsection{Summary}

To summarize, while the choice \(k(r) = \xi/r\) is convenient and robust for leading-order analysis, alternative scale identifications based on curvature invariants may better capture quantum gravity effects in strong-field regimes. Higher-order corrections are systematically calculable and provide further refinements to black hole quantum structure, potentially sharpening the conditions under which the Weak Gravity Conjecture is satisfied within Asymptotically Safe Quantum Gravity.

\end{document}